**Making neural networks understand internal heat transfer using Fourier-transformed thermal diffusion wave fields**

Pengfei Zhu[a], Hai Zhang[a,b,*], Clemente Ibarra-Castanedo[a], Xavier Maldague[a], Andreas Mandelis[c,d,*]

[a]Department of Electrical and Computer Engineering, Computer Vision and Systems Laboratory (CVSL), Laval University, Quebec G1V 0A6, Canada

[b]Centre for Composite Materials and Structures (CCMS), Harbin Institute of Technology, Harbin 150001, China

[c]Center for Advanced Diffusion-Wave and Photoacoustic Technologies (CADIPT), Department of Mechanical and Industrial Engineering, University of Toronto, Toronto, M5S 3G8, Canada

[d]Institute for Advanced Non-Destructive and Non-Invasive Diagnostic Technologies (IANDIT), Faculty of Applied Science and Engineering, University of Toronto, Toronto, M5S 1A4, Canada

[*]Corresponding author: hai.zhang.1@ulaval.ca (H.Z.), mandelis@mie.utoronto.ca (A.M.)

**Abstract**

Heat propagation is governed by phonon interactions and mathematically described by partial differential equations (PDEs), which link thermal transport to the intrinsic properties of materials. Conventional experimental techniques infer thermal responses based on surface emissions, limiting their ability to fully resolve subsurface structures and internal heat distribution. Additionally, existing thermal tomographic techniques can only shoot one frame from each layer. Physics-informed neural networks (PINNs) have recently emerged as powerful tools for solving inverse problems in heat transfer by integrating observational data with physical constraints. However, standard PINNs are primarily focused on fitting the given external temperature data, without explicit knowledge of the unknown internal temperature distribution. In this study, we introduce a Helmholtz-informed neural network (HINN) to predict internal temperature distributions without requiring internal measurements. The time-domain heat diffusion equation was converted to the frequency-domain and becomes the pseudo-Helmholtz equation. HINN embeds this pseudo-Helmholtz equation into the learning framework, leveraging both real and imaginary components of the thermal field. Finally, an inverse Fourier transform brings real-part and imagery-part back to the time-domain and can be used to map 3D thermal fields with interior defects. Furthermore, a truncated operation was conducted to improve computational efficiency, and the principle of conjugate symmetry was employed for repairing the discarded data. This approach significantly enhances predictive accuracy and computational efficiency. Our results demonstrate that HINN outperforms state-of-the-art PINNs and inverse heat solvers, offering a novel solution for non-invasive thermography in applications spanning materials science, biomedical diagnostics, and nondestructive evaluation.

## I. INTRODUCTION

Heat propagation is governed by random kinetic energy interactions between phonons. Subsequently, heat transfer theory is formulated by partial differential equations (PDEs), which not only quantifies the response of materials to heat flux but also reveals the intrinsic microscopic structure and physical properties of matter. Existing experimental radiation heat transfer techniques such as calorimetry, infrared camera imaging, and thermometry infer the thermal response by measuring the exterior surface heat radiation emitted by an object. According to the thermal perturbation data recorded by infrared (IR) sensors and parametrized using mathematical models, the thermophysical properties and subsurface structures can be inversely calculated. Therefore, numerous applications have been developed in industry and manufacturing[1-3], biomedicine[4,5], nondestructive testing and quality control[6-8], etc. However, quantifying heat transfer is constrained by the limited information available from the exterior boundaries' thermal response alone. A detailed knowledge of thermal diffusion-wave fields can be obtained if the temperature values of each node in the overall computational domain are known. Then it is possible to estimate the thermal diffusivity, energy distribution, boundary effects and localized material anomalies[9-12].

In the area of internal temperature prediction, Zhang *et al.*[13] addressed the temperature monitoring challenge in lithium-ion batteries for electric vehicles by developing an internal temperature predictive model based on the thermal network method. Patil *et al.*[14] presented a novel inverse analysis framework for predicting the internal temperature of cylindrical heat-generating bodies based on coolant temperature measurements. Wen *et al.*[15] introduced the Kalman smoothing (KS) technique, combining Kalman filtering and Rauch-Tung-Striebel smoothing to solve the inverse radiation-conduction heat transfer problem using future temperature measurements. Fan *et al.*[16] proposed a self-training feedforward neural network for predicting lithium-ion battery surface temperature 300 seconds in advance, enhancing the battery management system performance and safety. However, limitations of the above-mentioned methods highlight the need for further refinement, hybrid approaches, and improved data acquisition techniques for more accurate and robust internal temperature predictions.

Three-dimensional reconstruction based on thermal tomography can only generate one-frame thermal images at different depths/layers[17,18]. It is still far from inverting the distribution and variation of the entire internal temperature field. Using invasive methods such as inserting sensors into the body of an object will destroy its original structure. Therefore, it can be concluded that the major reconstruction issue comes from the lack of observation data, especially internal temperature data. In recent years, physics-informed neural networks (PINNs) have attracted significant attention in solid/fluid mechanics[19-21], quantum mechanics[22], heat transfer[23,24], electromagnetic fields[25], etc. By leveraging automatic differentiation to efficiently solve partial differential equations (PDEs), PINNs integrate observational data with physical constraints, enabling the handling of complex geometries and multiphysical process coupling problems without the need for explicit discretization. Moreover, their strong high-dimensional fitting capability makes them suitable for solving inverse problems and tackling scenarios with sparse data, providing an efficient and flexible solution for complex system modeling where traditional numerical methods fall short. However, we found that the neural networks still failed to learn the internal features of heat transfer even if using PINNs. There are three main reasons for that: First, only an external temperature point cloud is used. Second, the problem is ill-posed with unknown thermal boundary conditions. Last but not least, general heat transfer problems are time dependent, not steady state. Transient heat transfer is a parabolic PDE compared with the elliptic PDE character of steady heat transfer problems, which makes the problem much harder to solve due to the time-dependent term. Furthermore, complicated boundary conditions (such as thermal shocks) and high computational cost also make the network ineffective.

The prediction of the internal temperature field can be treated as solving an ill-posed problem, i.e., lacking internal temperature data and boundary conditions. Here, a Helmholtz-informed neural network (HINN) was designed to predict the internal temperature distribution and variation of a solid object. By feeding all discrete data at the external surface of the object into the HINN, it is possible to construct an inverse problem solver based on PINNs. The time-domain heat diffusion equation was converted to the frequency-domain and becomes the pseudo-Helmholtz equation[26]

$$\left(\frac{\partial^2}{\partial x^2} + \frac{\partial^2}{\partial y^2} + \frac{\partial^2}{\partial z^2} - \frac{i\omega}{\alpha}\right)\hat{T} = 0 \qquad (1)$$

was embedded into machine learning instead of the time-domain heat conduction equation. The pseudo-Helmholtz equation-based training dataset includes both real and imaginary parts. Finally, an inverse Fourier transform brings real-part and imagery-part back to the time-domain and can be used to map 3D thermal fields with interior defects. Furthermore, a truncated operation was conducted to improve computational efficiency, and the principle of conjugate symmetry was employed for repairing the discarded data. After a very short training period, the HINN was found to be able to predict the overall temperature distribution and variations.

**II. THEORY**

The physics-informed neural networks simulate the core idea of Green's function. Consider the heat equation on a bounded domain $\Omega \subset \mathbb{R}^d$ ($d$-dimensional Euclidean space) with boundary $\partial\Omega$ over a time interval $t \in [0, \xi]$:

$$\frac{\partial u}{\partial t}(x,t) = \alpha \nabla^2 u(x,t) \tag{2}$$

with

$$u(x,0) = u_0(x), \qquad u|_{\partial\Omega} = g(x,t) \tag{3}$$

where $\alpha$ is the thermal diffusivity, $u_0$ is the initial temperature, and $g(x,t)$ is the boundary temperature. Let $G(x,t;y,\tau)$ denote the Green's function (heat kernel) for the homogeneous heat equation in $\mathbb{R}^d$, given by:

$$G(x,t;y,\tau) = \frac{1}{(4\pi\alpha(t-\tau))^{d/2}} e^{\left(-\frac{\|x-y\|^2}{4\alpha(t-\tau)}\right)}, \qquad \text{for } t > \tau \tag{4}$$

Then the solution to the initial-boundary value problem can be represented as:

$$u(x,t) = \int_\Omega G(x,t;y,0)u_0(y)dV_y + \int_0^t \int_{\partial\Omega}[G(x,t;y,\tau)\frac{\partial u}{\partial n}(y,\tau) - \frac{\partial G}{\partial n_y}(x,t;y,\tau)g(y,\tau)]dS_y d\tau \quad (5)$$

This formula illustrates that the temperature field $u(x,t)$ inside the domain is entirely determined by the initial condition $u_0$ and the boundary data $g$. Given both $u_0$ and $g$, one can solve a boundary integral equation for $\frac{\partial u}{\partial n}$.

In the PINNs framework, a trial function $u_\theta(x,t)$ is introduced, parameterized by neural network weights $\theta$, and trained by minimizing a composite loss function:

$$\mathcal{L}(\theta) = \int_0^\xi \int_\Omega |\frac{\partial u_\theta(x,t)}{\partial t} - \nabla^2 u_\theta(x,t)|^2 dV dt + \int_\Omega |u_\theta(x,0) - u_0(x)|^2 dV + \int_0^\xi \int_{\partial\Omega} |u_\theta(x,t) - g(x,t)|^2 dS dt \quad (6)$$

where the first term $\mathcal{L}_R$ on the right hand side of Eq. (6) forces the interior residual of the heat equation, and the second term $\mathcal{L}_{IC}$ enforces the initial condition, the third term $\mathcal{L}_{BC}$ enforces the boundary condition. According to the well-posedness theory of the heat equation, any sufficiently smooth function $u_\theta(x,t)$ that minimizes all three terms to zero must coincide with the true solution $u(x,t)$. This reflects the same principle as the Green's function representation: the solution is fully governed by its initial and boundary conditions.

## A. Physics-informed neural networks

The principal challenge in the prediction of internal temperature fields has been introduced in the previous section. It is difficult to solve this problem because it is an ill-posed inverse problem with unknown boundary conditions and internal temperature information. Fortunately, a few data sets of the surrounding surface temperature are available and can be used to infer the subsurface heat transport including at internal discontinuities / defects which create local thermal resistance that affects the surface temperature. The difficulty of modelling is extremely high, so investigations are usually simplified considering one-dimensional heat transfer problems instead of full 3-D thermal fields. Unlike light and sound fields, initial perturbations in heat conduction quickly dissipate, making the inverse problem ill-posed and difficult to reconstruct earlier thermal distributions from later observations. Additionally, in practical scenarios, measuring exact thermal boundary conditions is challenging, such as associated with commonly used flash lamp heating, linear scanning heating or chirp pulse heating. Therefore, an accurate and efficient inverse problem solver is needed for exploring the internal thermal response of condensed matter.

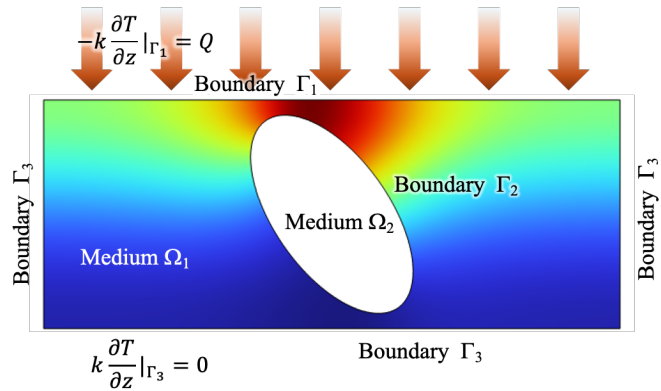

**FIG. 1.** Ill-posed inverse problem in thermal tomography. Boundary $\Gamma_1$ is the front surface. Boundary $\Gamma_2$ is the interface between medium $\Gamma_1$ and $\Gamma_2$. Boundary $\Gamma_3$ contains side and back boundaries. Generally,

boundary $\Gamma_3$ is subject to the Robin boundary condition (see Eq. (11)). The target is to inversely calculate the time-dependent temperature distribution in medium $\Gamma_1$ and $\Gamma_2$ through pre-known data on $\Gamma_1$ and boundary $\Gamma_3$.

The general geometry of the ill-posed heat transfer inverse problem is shown in Fig. 1. There are three kinds of boundaries outside the body of the sample, labeled $\Gamma_1$, $\Gamma_2$, and $\Gamma_3$. The boundary $\Gamma_1$ is subject to an incident heat flux with power $Q$, while the boundary $\Gamma_3$ is subject to a convective boundary condition. The boundary $\Gamma_2$ is unknown and reflects the inner interface discontinuity. $Q$ is not always a constant value or a well-defined function. This is also a significant difference between the method proposed here and other reports[27-29]. Generally, in pulsed thermography, incident heat flux is considered as a Dirac impulse even a single square or rectangular wave, which is only an approximation. In the case of a single square wave, the real relaxation signal in the frequency domain has a finite effective bandwidth, with high-frequency components gradually attenuating. While the Fourier series of an ideal square wave theoretically contains infinite harmonics, implying infinite bandwidth, in practical systems high-frequency components are limited by physical and instrumentation constraints, resulting in a finite effective bandwidth. For the square-wave case, information about incident heat sources is no longer needed since the neural networks can learn the heat source features according to the temperature variation and distribution at the front (incident) side. In PINNs, random discrete sampling is performed for the training datasets. Then the sampling data are fed into neural networks for training. However, specific pixels in the acquired data may be sensitive to the internal thermal impedances due to defects or voids. In this case, random discrete sampling is discarded, and the neural network needs to capture the data from all pixels at each time unit. After training the PINNs, node information was fed into the network for predicting the internal temperature field distribution and variation (Fig. 2(a)). Furthermore, the material thermal diffusivity is unknown in most cases. In this work, after assuming an educated initial guesstimate, the thermal diffusivity is set as a trainable parameter in the neural network. The structure of PINNs in the time domain is shown in Fig. 2(a).

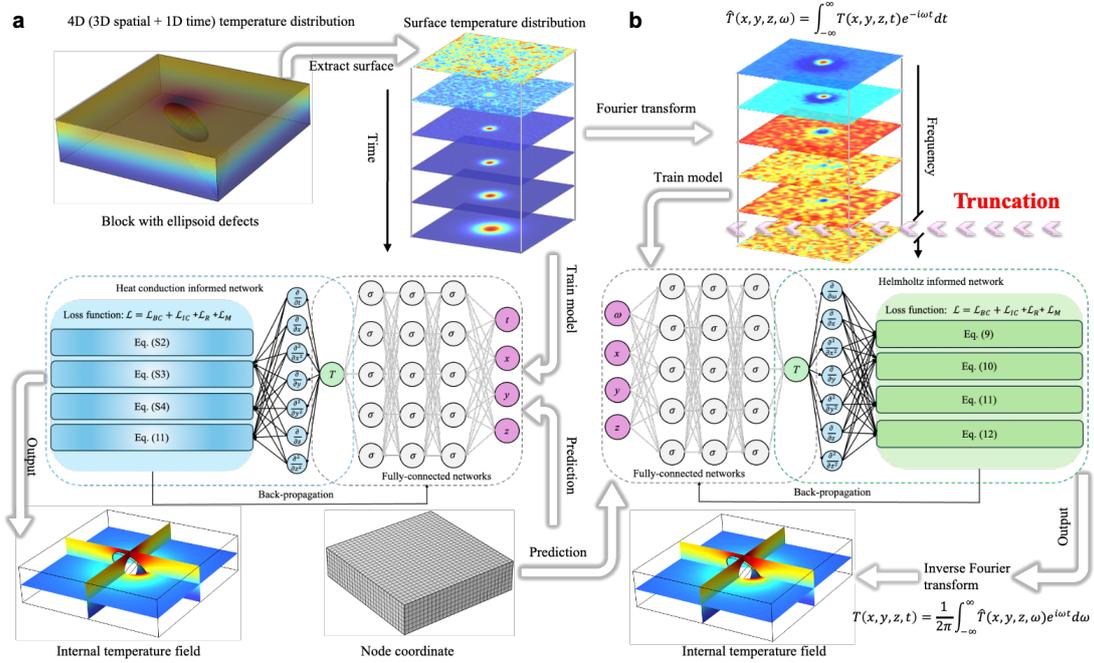

**FIG. 2.** Schematic illustration of PINNs and HINN. (a) The surface temperature data are extracted and fed into the PINNs. Internal temperature field (4D) can be predicted by feeding the node coordinates into the trained PINNs. (b) The Fourier transforms of surface temperature data are fed into HINN where heat conduction informed network is converted to Helmholtz informed network. Truncation is then performed to reduce the computation time.

**B. Helmholtz-informed neural network**

After designing the PINNs in the time domain, there is a challenge that needs to be solved, i.e., the computation time. For a commonly used infrared camera, the spatial resolution is lower than about 1000 × 1000 pixels (which can be adjusted artificially). The time resolution depends on the IR camera frame rate. In industrial monitoring[30], the frame rate is often set to 50 Hz, while it should be as high as possible (> 100 Hz) in biomedical imaging[31] (since it requires higher depth resolution). This is due to the fact that frequency is directly related to depth resolution. In the case of industrial inspections, a test typically lasts 14 s. Therefore, the size of training data at 50 Hz will be 1000 × 1000 × (14 × 50). For PINNs, processing new samples means retraining. Such a large amount of data and retraining requirements are unacceptable in practice. Here, a Helmholtz-informed neural network (HINN) is proposed to alleviate the problem of computation time. First, the heat transfer problem can be formulated as

$$\frac{\partial T_{1,2}}{\partial t} = \alpha_{1,2}(\frac{\partial^2 T_{1,2}}{\partial x^2} + \frac{\partial^2 T_{1,2}}{\partial y^2} + \frac{\partial^2 T_{1,2}}{\partial z^2}) \tag{7}$$

with boundary conditions (Fig. 1)

$$-k_1 \frac{\partial T_1}{\partial z}|_{\Gamma_1} + Q(x, y, t) = 0 \tag{8}$$

$$T_1|_{\Gamma_2} - T_2|_{\Gamma_2} = 0 \tag{9}$$

$$k_1 \frac{\partial T_1}{\partial z}|_{\Gamma_2} - k_2 \frac{\partial T_2}{\partial z}|_{\Gamma_2} = 0 \tag{10}$$

$$-k_1 \frac{\partial T_1}{\partial z}|_{\Gamma_3} - h(T_1 - T_\infty) = 0 \tag{11}$$

where $T_{1,2}$ is the temperature of medium $\Omega_1$ and $\Omega_2$, respectively, $k_1$ and $k_2$ are the thermal conductivities of media $\Omega_1$ and $\Omega_2$, $T_\infty$ is the ambient temperature, $t$ is time, $\alpha_{1,2}$ is thermal diffusivity of media $\Omega_1$ and $\Omega_2$, respectively. $(x, y, z)$ are spatial coordinates along $x, y, z$ direction. It is noted that boundary conditions (9), (10), and even (11) are unknown. Taking the Fourier transform of the time-dependent heat diffusion equation Eq. (7) for the temperature $T(x, y, z, t)$

$$\hat{T}(x, y, z, \omega) = \int_{-\infty}^{\infty} T(x, y, z, t)e^{-i\omega t}dt \qquad (12)$$

yields the pseudo-Helmholtz Eq. (1). The frequency domain Fourier transform $\hat{T}(x, y, z, \omega)$ is the thermal-wave field (spectrum) and $\omega$ is the spectral angular frequency. The structure of HINN is shown in Fig. 2(b). It should be mentioned that the amount of data and computation time cannot be reduced working in the frequency domain. The advantage of HINN is that effective information is focused on the low frequency range. Therefore, the data of each pixel can be truncated at a certain frequency ($1/n*f_s$, where $f_s$ is the sampling frequency and $n$ is an operator-controlled value). It should be mentioned that the sampling rate must minimally satisfy the Nyquist sampling theorem so that spectral aliasing can be avoided. Because the thermal diffusion length is directly related to the frequency and thermal diffusivity ($\mu = \sqrt{2\alpha/\omega}$) where $\omega = 2\pi f$. From the viewpoint of time-domain signal analysis, this process is tantamount to low-pass filtering, which has benefits in denoising and enhancing signal-to-noise ratio (SNR). To visualize the depth information related to frequency, we consider five commonly used materials: ceramics, metals, basalt, carbon, and wood. It is possible to observe the thermal diffusion length varying with the frequency (Fig. 3). Table I shows the details of different material thermophysical properties. Due to the inverse square root relationship between the thermal diffusion length and frequency, the penetration depth of a thermal wave decays extremely rapidly with increasing frequency, especially for samples with low thermal diffusivity. Fortunately, the thermal wave can still reach a very near-surface depth in the low frequency range which is truncated at $f_t \leq \frac{1}{10} f_s$ for samples with low thermal diffusivity. This means that there is no need to consider the high-frequency range $f > \frac{1}{10} f_s$. In contrast, for samples with high thermal diffusivity, the penetration amplitude of a thermal wave decays at a slower depth rate and higher frequencies are required to reach the near-surface region. Therefore, one needs to make sure the truncation frequency ($f_t \leq \frac{1}{10} f_s$) can encompass as much information about the subsurface structure as possible. In finite element analysis it is possible to observe the thermal distribution of a metallic plate at different frequencies (Fig. 3). When the truncation frequency is set at 5 Hz, it can contain most of the effective information about the subsurface structure. Compared with HINN, the reason why the PINNs cannot be truncated in the time domain is that the penetration depth increases with time, $z = 2\sqrt{\alpha t}/\sqrt{\pi}$[34]. If truncation occurs at early times the result is loss of information from deep layers.

TABLE I. Thermophysical properties of five solid state samples.

| Material | Thermal conductivity (W·m$^{-1}$·K$^{-1}$) | Density (kg·m$^{-3}$) | Specific heat (J·K$^{-1}$·kg$^{-1}$) | Thermal diffusivity (m$^2$·s$^{-1}$) |
|---|---|---|---|---|
| Ceramic | 18 | 3800 | 750 | 6.32×10$^{-6}$ |
| Metal | 16.2 | 7900 | 477 | 4.30×10$^{-6}$ |
| Basalt | 3 | 2600 | 790 | 1.46×10$^{-6}$ |
| Carbon | 0.5 | 1400 | 800 | 4.46×10$^{-7}$ |
| Wood | 0.12 | 400 | 1600 | 1.88×10$^{-7}$ |

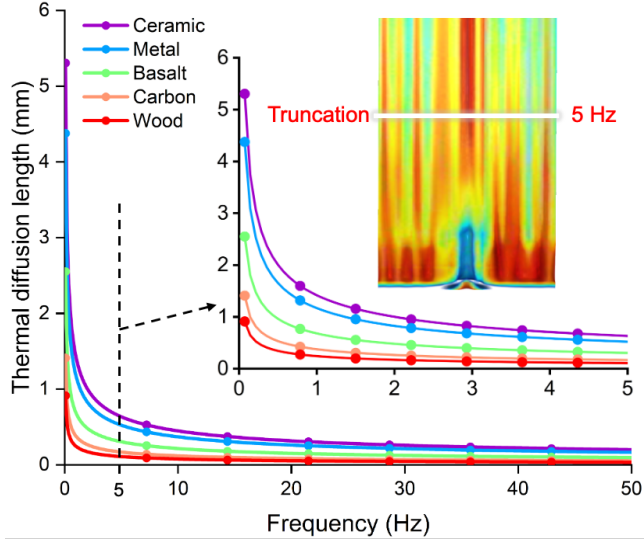

**FIG. 3.** Analysis of truncation frequency. Five types of materials (ceramic, metal, basalt, carbon, and wood) were used to calculate the relationship between the thermal diffusion length and frequency. The cross-sectional image presents the thermal wave varying with frequency for the middle section of the metallic sample.

The loss function is used to quantify the difference between the networks' predictions and the actual data. The goal of network training is to minimize this loss and improve the networks' prediction accuracy so that the predictions align more closely with the true values. For HINN, the loss function is given by:

$$\mathcal{L} = \mathcal{L}_{BC} + \mathcal{L}_{IC} + \mathcal{L}_R \tag{13}$$

$$\mathcal{L}_{BC} = \frac{1}{N_{BC}} \sum_{i=1}^{N_{BC}} |\hat{\mathbf{u}}(\mathbf{x}^i, \omega^i) - \hat{\mathbf{u}}_{data}^i|^2 \tag{14}$$

$$\mathcal{L}_{IC} = \frac{1}{N_{IC}} \sum_{i=1}^{N_{IC}} |i\omega\hat{\mathbf{u}}(\mathbf{x}^i, \omega^i) - T_{ini} - \nabla^2\hat{\mathbf{u}}(\mathbf{x}^i, \omega^i)|^2 \tag{15}$$

$$\mathcal{L}_R = \frac{1}{N_R} \sum_{i=1}^{N_R} |(\nabla^2 + \frac{i\omega}{\alpha})\hat{\mathbf{u}}(\mathbf{x}^i, \omega^i)|^2 \tag{16}$$

where $\mathcal{L}_{BC}$, $\mathcal{L}_{IC}$, $\mathcal{L}_R$ penalize the residuals, that is, the difference between theoretically correct values and network predicted values of the boundary conditions, initial conditions, and governing equations, respectively. $N_{BC}$, $N_{IC}$, and $N_R$ are the numbers of data points for different terms. Furthermore, there is an additional loss function in PINNs, which forces the output response to match the response of the heat conduction equation under investigation:

$$\mathcal{L}_M = \frac{1}{N_M}\sum_{i=1}^{N_M} |\hat{\mathbf{u}}(\mathbf{x}^i, \omega^i) - \hat{\mathbf{u}}_{data}^i|^2 \qquad (17)$$

This type of computation is achieved in the PINNs framework using automatic differentiation[32], an operation which is a key enabler for PINNs. It combines the derivatives of the constituent operations using the chain rule and outputs the derivative of the overall composition, defined as the entire sequence of mathematical operations that constitute the neural network's forward propagation. The normalized process was performed on all training datasets (only on the thermal-wave values instead of spatial coordinates). All models (PINNs and HINN) were implemented with the PyTorch framework, a code database in Python software, specifically developed for achieving deep learning, and then trained using NVIDIA 4060 Titan GPUs[33,34].

### III. Application to heat transfer with unknown internal boundary

### A. Problem description: prediction failure in PINNs

After training 10,000 epochs ("epoch" refers to one complete propagation through the entire training dataset), it is possible to predict the thermal-wave field based on the pre-known node coordinate, as shown in Fig. 4. In this section, all simulation modelling is based on the metallic plate with a rectangular void of Fig. 4. The material properties are shown in Table I. It is clear that the absolute error between the predicted temperature and real temperature decreases with time. However, The PINNs can only fit and approximate the temperature values from given boundary nodes. For internal thermal distribution, especially the thermal resistance effect at discontinuity interfaces, PINNs cannot provide an accurate response. A similar problem has been identified elsewhere[23], as feeding into the temperature values from internal nodes it is difficult to make PINNs learn the mechanism of internal boundary localized heat transfer. On the contrary, PINNs can effectively predict the internal deformation according to the only deformation data from outside node and boundary conditions[20]. The main reason is that deformation is typically governed by elasticity equations, which are elliptic PDEs. Those functions enforce a strong global coupling, meaning that information from external boundaries can propagate effectively to the interior domain. In statics (mechanics), deformation follows the principle of minimum potential energy, making it easier to infer internal displacement from boundary conditions. Furthermore, compared with elasticity equations, transient heat conduction is time-dependent and the temperature variation depends on the thermophysical properties of materials.

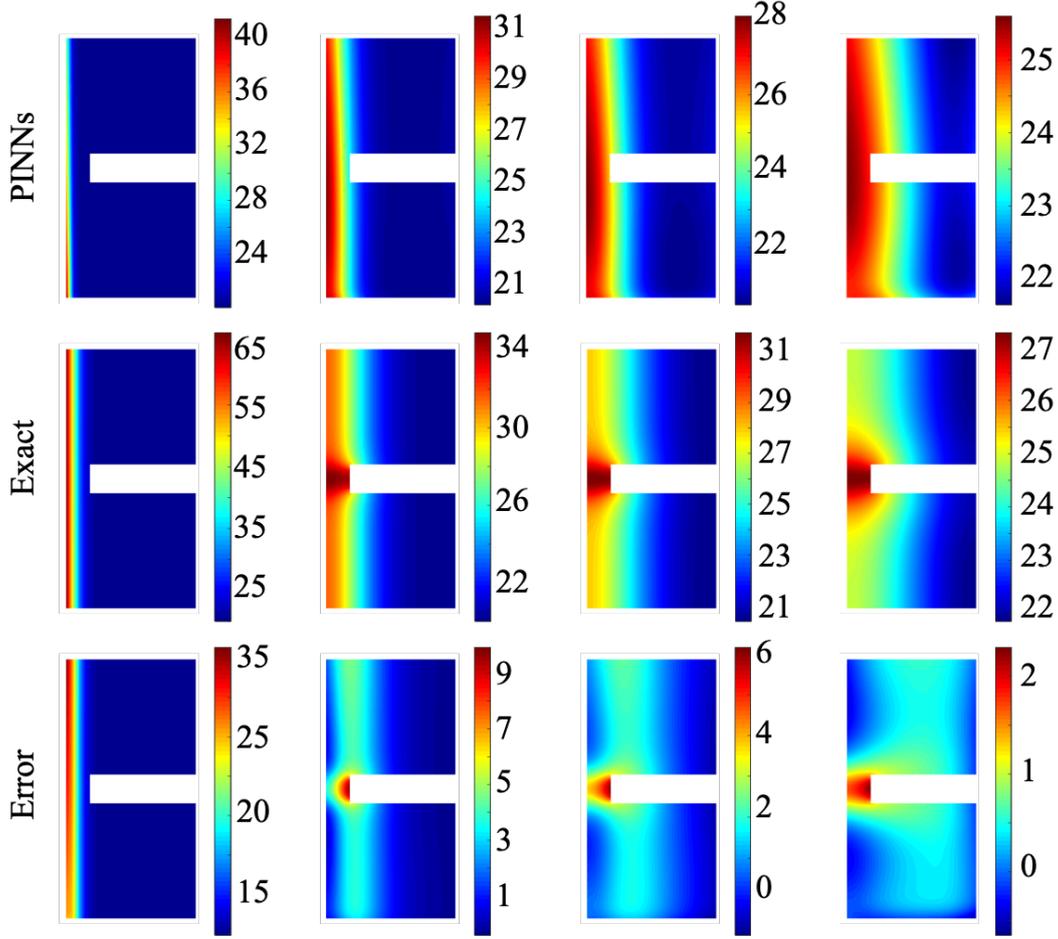

**FIG. 4.** Analysis of simulation data using PINNs. The first row is the predicted temperature field based on PINNs at 0.02, 0.18, 0.28, 0.98 s. The second row is the exact temperature field, calculated using finite element method (FEM) simulations. The third row is the absolute error between the predicted temperature field (first row) and the exact temperature field (second row). The units are °C.

### B. Helmholtz-informed neural network: internal thermal field prediction

Different from PINNs, the input data in HINN contains real and imaginary parts. Therefore, the loss function in Eqs. (14)-(17) can be divided into real and imaginary loss functions. In this case, HINN directly optimizes the loss function in the frequency domain. By applying the Fourier transformation, differential operators are converted into algebraic operations, making high-order derivatives more stable. Only a few sampling points are needed in the frequency domain to obtain accurate predictions.

HINN was used to predict the internal temperature distribution, as shown in Fig. 5. Unlike conventional PINNs, HINN can effectively learn the physical diffusive behavior of heat transfer. For instance, there is a strong heat accumulation at the discontinuous interface of Fig. 5(a). It is possible to calculate the absolute error between HINN results and exact results. Compared with PINNs (Fig. 4), it is clear that the absolute error in HINN becomes lower.

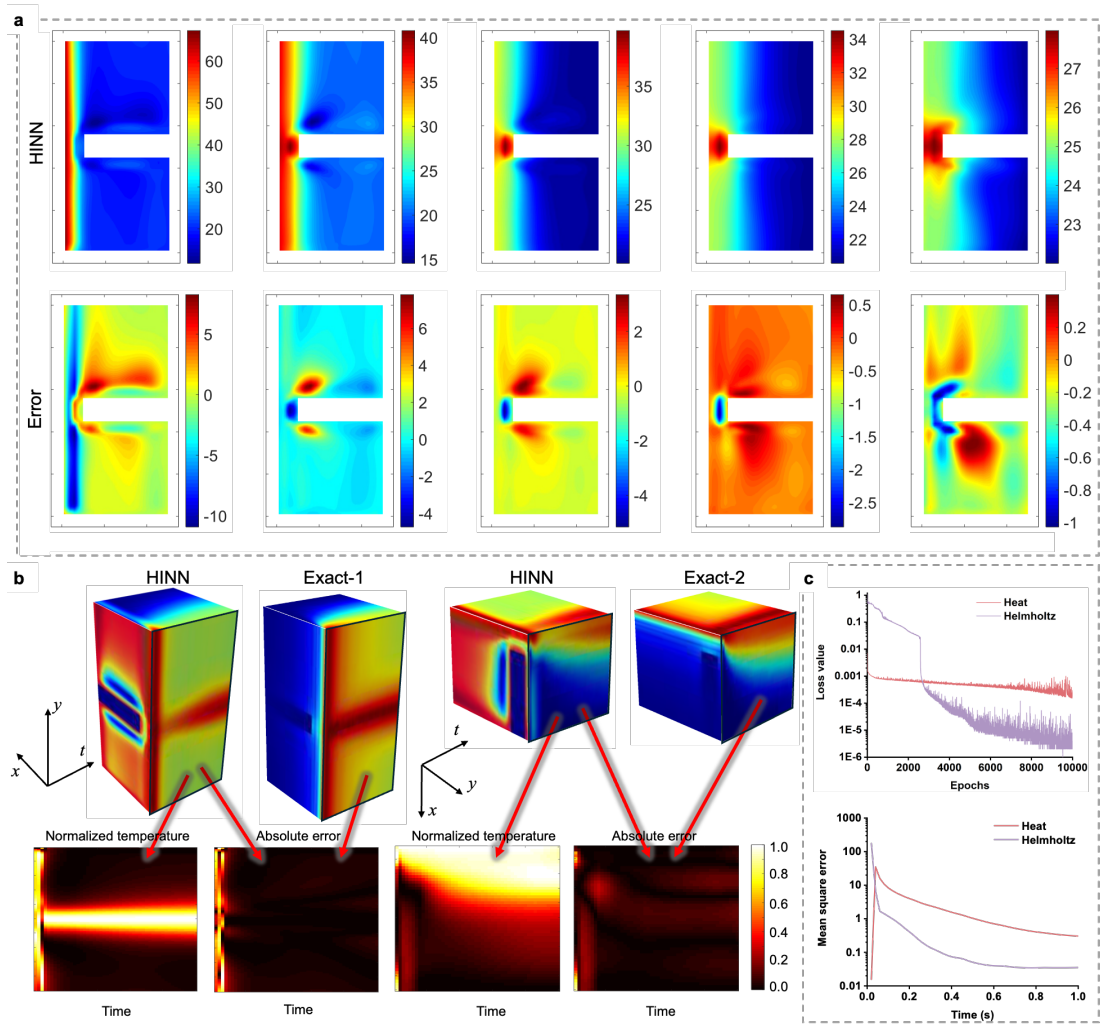

**FIG. 5.** Analysis of simulation data using HINN. (a) The first row is the predicted temperature field based on HINNs at 0.02, 0.08, 0.18, 0.28, 0.98 s. The second row is the absolute error. (b) Three-dimensional time-domain thermal distribution of predicted results, exact results, and internal temperature distribution. (c) Quantitative comparison between original PINNs and HINN. The temperature units are °C, and the time units are s (seconds).

To further observe the 3D thermal distribution, we visualized the 3D temperature field and cross-section images of HINN and exact results, as shown in Fig. 5(b). The thermal distributions of HINN and exact results are almost identical. Similarly, thermal distribution images were selected from two side views (Fig. 5(b)). To better observe the heat distribution along the time coordinate, a normalization process was applied at each time increment (seconds). The absolute error images in Fig. 5(a) display higher accuracy (maximum error is less than 1) for the HINN method than that of the PINNs in Fig. 4 (maximum error is more than 2). Finally, a quantitative comparison between original PINNs and HINN was implemented, as shown in Fig. 5(c). It is possible to observe that the loss value of HINN at 10,000 epochs is two orders of magnitude lower than that of the original PINNs. The lower loss indicates that the predicted temperature values are much closer to the real ones. Furthermore, the mean square error of HINN is also significantly lower than that of PINNs.

**B. Reducing computation cost: integrating with conjugate symmetry**

To validate the truncation effect as depicted in Fig. 2(b), only the first (lowest) ten frequency components were selected from the entire spectrum according to the criterion $f_t \leq \frac{1}{10} f_s$. Then HINN was employed to train these finite datasets and predict the transient temperature field distribution based on the inverse Fourier transform

$$T(x,y,z,t) = \frac{1}{2\pi} \int_{-\infty}^{\infty} \hat{T}(x,y,z,\omega) e^{i\omega t} d\omega \qquad (18)$$

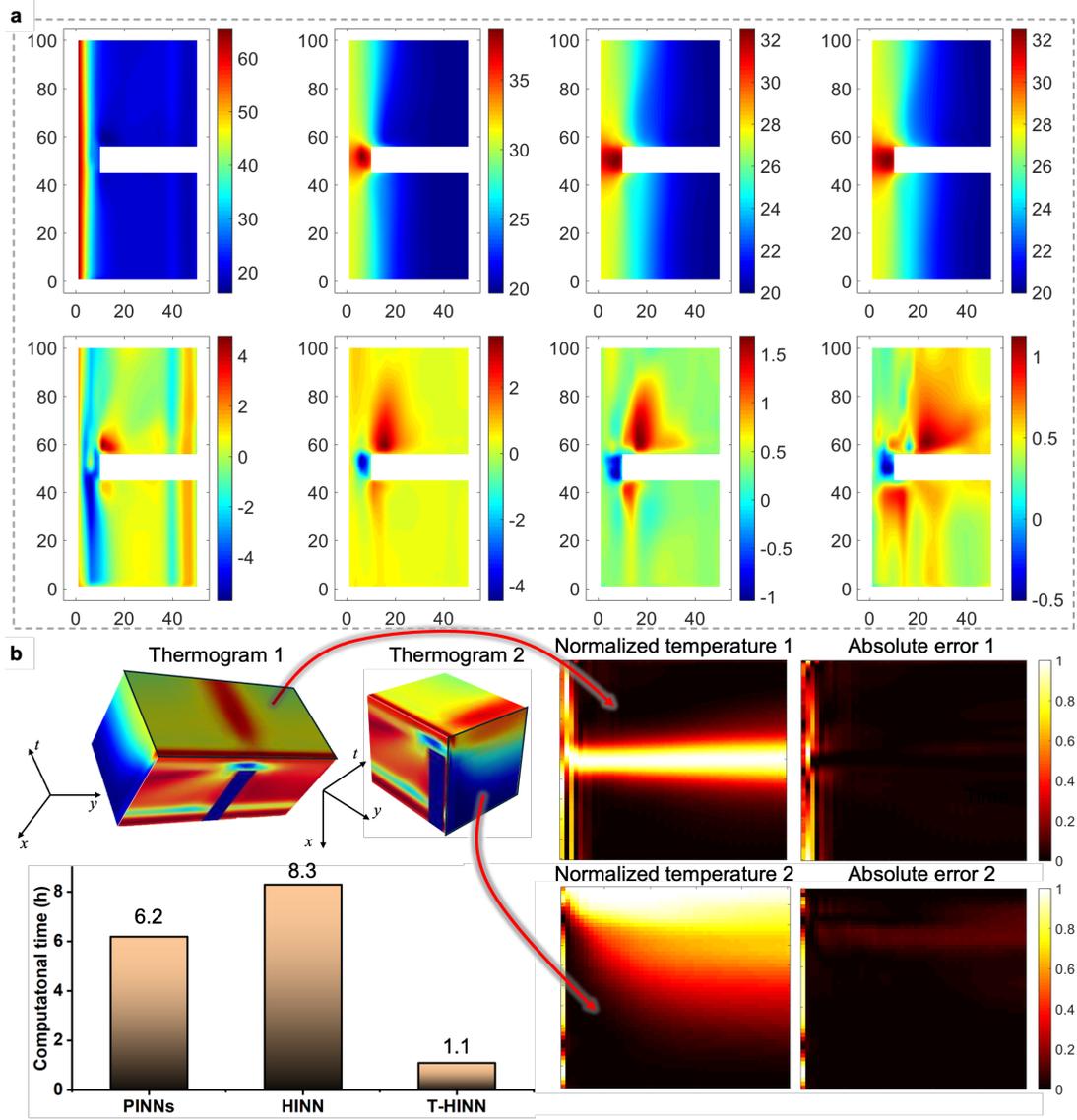

**FIG. 6.** Analysis of truncation effects in HINN. (a) The first row is the predicted temperature field based on PINNs at 0.02, 0.18, 0.28, 0.98 s. The second row is the absolute error. (b) Three-dimensional thermal distribution of predicted results and exact (simulated) results and internal temperature distribution. T-HINN denotes the truncated HINN. The units are °C.

To more accurately restore the original temperature values, conjugate symmetry was used, as the original temperature data consist of real-valued signals. Conjugate symmetry means that for every positive frequency component of a real temperature signal, there is a corresponding negative frequency component representing its complex conjugate, ensuring that the inverse Fourier transform accurately reconstructs real-valued temperature data. The 11th and up to the (*N*-10)-th frequency components were filled with their corresponding values consistently with the Nyquist sampling theorem. Here, *N* is the length of the spectrum, while frequency values outside this range were set to zero.

Figure 6 shows the analysis of the truncation effect in HINN using the same metal sample introduced in Fig. 2, with thermophysical properties shown in Table I. It was possible to find that the truncated HINN (T-HINN) can also learn the heat transfer kinetics at the discontinuous boundary (void area). Strong thermal impedance was thus predicted as shown in Fig. 6(a). The absolute error images are in the second row of Fig. 6(a). Compared with results in HINN, T-HINN exhibits almost the same absolute error values. The 3D thermograms are shown in Fig. 6(b). As shown in Fig. 6(b), the normalized temperatures 1 and 2 and absolute errors 1 and 2 were calculated according to thermograms 1, 2 and the exact thermogram (Exact-1, Exact-2, Fig. 5(b)). The absolute errors 1 and 2 are very small, which means that the thermal distributions along time and space are almost the same as the exact distribution. Finally, the computation time of PINNs, HINN, and T-HINN were compared. Although frequency-domain methods require handling complex signals (including real and imaginary parts) and additional loss functions, which seem to increase computational complexity compared to PINNs, the phase information in these complex signals is actually equivalent to time delays in the time domain: $\mathcal{F}\{f(t-\tau)\} = F(\omega)e^{-i\omega\tau}$, where $F(\omega)$ is the Fourier transform of the original signal $f(t)$ and $\tau$ is the time delay. In other words, the phase in the frequency domain carries the same essential information as time delays, which are crucial for accurately describing the dynamic characteristics of the signal. Therefore, while frequency-domain processing may appear to add computational overhead, selecting only key frequency components for training can effectively reduce computation time while preserving essential physical information, thus balancing computational efficiency and accuracy.

## IV. EXPERIMENTAL VALIDATION

A specimen (carbon fiber-reinforced polymer, CFRP) was used to test the proposed technique, as shown in Fig. 7. The specimen consisted of a 10-ply carbon fiber-reinforced polymer with 25 Teflon square inserts located at different depths (0.2 mm < z < 1.0 mm) and with different lateral sizes (3 mm < D < 15 mm).

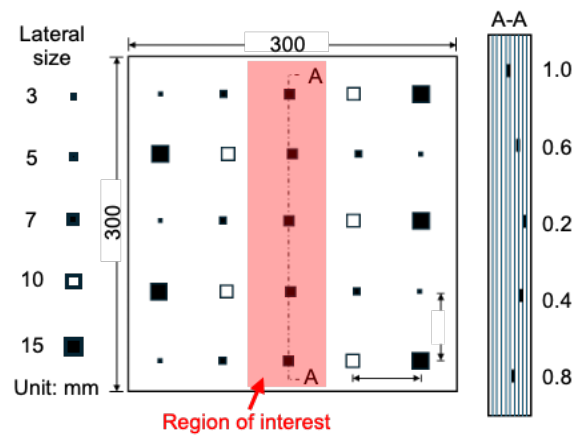

**FIG. 7.** Schematic image of the sample.

The schematic image of this work is shown in Fig. 8. The infrared thermography system utilized pulsed thermography (PT) in reflection and transmission mode, incorporating a cooled infrared camera (FLIR X8501sc, 3–5 μm, InSb, NEdT < 20 mK, 1280 × 1024 pixels) along with two xenon flash lamps (Balcar, 6.4 kJ each, 2 ms @

FWHM). The frame rate of the infrared camera was ~45 frames/s.

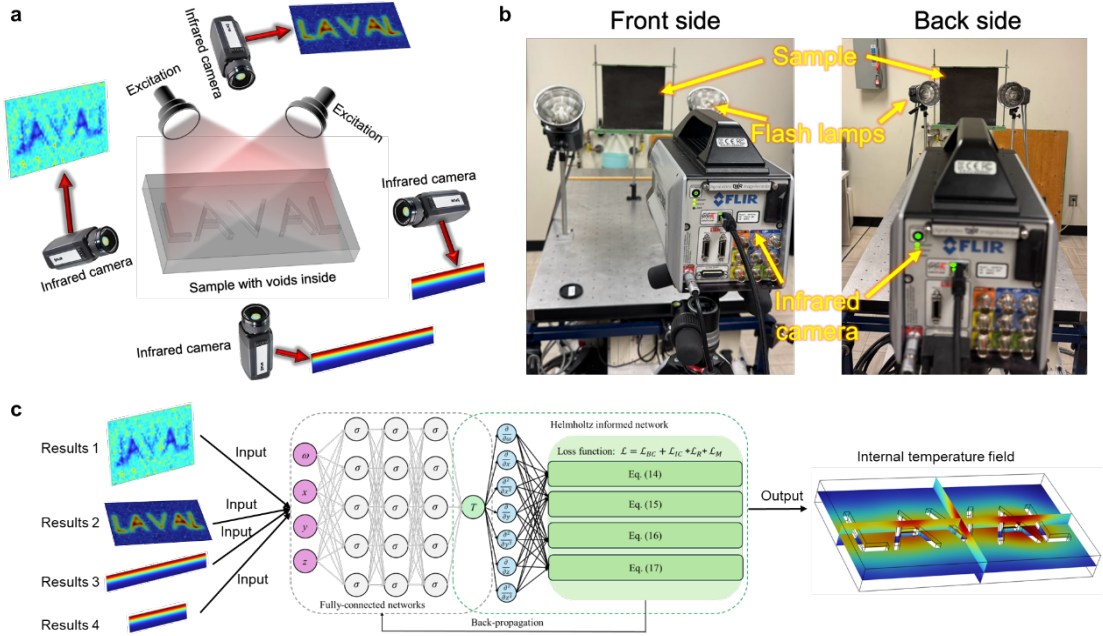

**FIG. 8.** Schematic image of this work. (a) Schematic image of experimental setup. (b) Photographs of experimental setup. (c) Schematic image of the prediction of internal (entire) temperature field.

Most PINNs studies are focused on simulation without experimental validations[35-37]. Here, we applied the proposed T-HINN to infrared thermographic experiments. The details of sample and experimental system were introduced in Section II. Different from simulation, it is an experimental fact that an infrared camera will generate noise during the image recording process. In order to reduce the effect of camera noise, the thermal signal reconstruction (TSR) method was employed in this work[38], as shown in Fig. 9(a). The pixels at the defect and sound areas were chosen and Fourier transformation of the transient data was performed to obtain the real and imaginary parts of the thermal wave spectrum[39]. The phase reflects the excellent denoising capability of the thermal signal reconstruction (TSR) technique, as shown in Fig. 9(a). Then the raw and TSR data were fed into the T-HINN model. It was observed that the loss value can decrease after TSR processing, as shown in Fig. 9(b). Due to the large aspect ratio (~150) of the tested sample, only the pixels around the middle defect area (0.2 mm defect depth) were chosen to visualize. Furthermore, the simulation was used to validate the right distribution of temperature fields, as shown in Fig. 9(c). It is obvious that T-HINN yields predictions consistent with the simulation (exact) results.

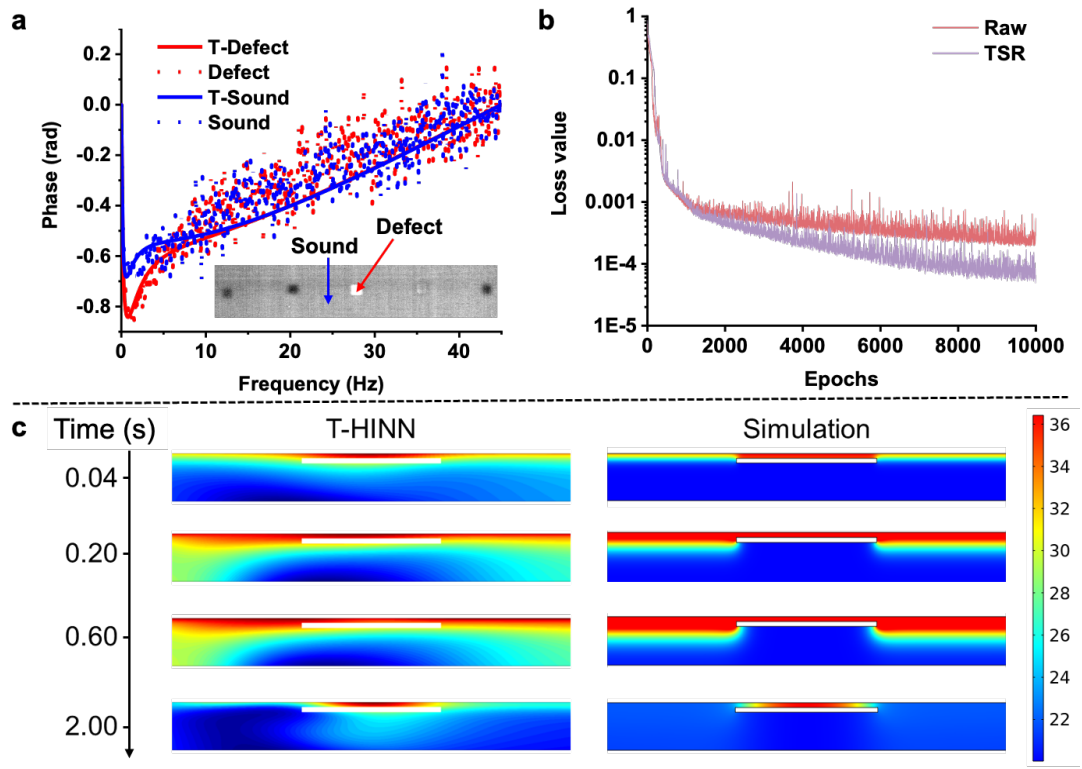

FIG. 9. Experimental prediction based on T-HINN. (a) Phase curves of original data and TSR data. (b) Loss curves of raw data and TSR data. (c) Internal temperature distribution prediction based on T-HINN and simulation. The units are °C.

## V. CONCLUSIONS

The heat transfer inside the body of a solid material is an unknown ill-posed inverse problem with unknown boundary conditions. Advanced inverse problem solvers including neural networks or mathematical modeling cannot provide an effective solution. For instance, the neural networks are typically limited to minimizing the temperature difference between the given data and the predicted values at corresponding locations. In other words, the neural networks primarily focus on fitting the observed external temperature data, without explicitly accounting for the unknown internal temperature distribution. In this work, a HINN was employed to predict the internal time-dependent temperature distribution without requiring any internal measurement. The time-domain heat diffusion equation was converted to the frequency-domain and became the pseudo-Helmholtz equation. HINN embedded this pseudo-Helmholtz equation into the learning framework, leveraging both real and imaginary components of the thermal field. Finally, an inverse Fourier transform brought real and imaginary parts back to the time-domain and was used to map 3D thermal fields with interior defects. It was found that the Helmholtz thermal-wave equation can make neural networks understand the characteristic of heat transfer, especially effects due to thermal impedance at interior interfaces representing defects or discontinuities. However, PINNs lack generalization abilities[40-44]. To address this issue, PINNs were considered as a computationally intensive one-time solver. To reduce the computation time, a truncated Helmholtz-informed neural network (T-HINN) was proposed. A truncated operation was conducted to improve computational efficiency, and the principle of conjugate symmetry was employed for repairing the discarded data. The network was found to significantly reduce the computation time and effectively predict and restore the internal temperature information. All simulations and experimental results demonstrated the excellent capability of the proposed HINN and T-HINN to precisely predict unknown internal temperature values and their variation with time. As mentioned in the Introduction, the prediction of internal temperature fields is not limited to industrial applications. It is also important in biomedical (early-stage disease thermal ablation and cryotherapy, inspection of blood flow and metabolism, tissue engineering and organ transplantation) and energy fields (inspection of battery heat generation, fuel cells, and solar panels, inspection of energy storage). In this study, we successfully applied this HINN (and T-HINN) to industrial inspection. One of the most challenging problems - internal thermal resistance - has been effectively addressed based on limited external observation points. These results demonstrate the potential of HINN (T-HINN) for broader applications in biomedical, industrial, and energy-related fields.

**CrediT authorship contribution statement**

**Pengfei Zhu:** Writing – original draft, Methodology, Investigation, Formal analysis, Data curation, Conceptualization. **Hai Zhang:** Writing – review & editing, Investigation, Project administration, Supervision. **Stefano Sfarra:** Writing – review & editing, Investigation. **Clemente Ibarra-Castanedo:** Resources, Data curation, Investigation. **Xavier Maldague:** Supervision, Project administration, Funding acquisition. **Andreas Mandelis:** Writing – review & editing, Methodology, Investigation.

**Declaration of competing interest**

The authors declare that they have no known competing financial interests or personal relationships that could have appeared to influence the work reported in this paper.

**Acknowledgements**

This work was supported by the Natural Sciences and Engineering Research Council (NSERC) Canada through the CREATE 'oN DuTy!' program (Grant no. 496439-2017) and the Canada Research Chair in Multipolar Infrared Vision (MiViM). AM gratefully acknowledges the Natural Sciences and Engineering Research Council of Canada (NSERC) Discovery Grants Program (RGPIN-2020-04595), the Canada Foundation for Innovation (CFI) Research Chairs Program (950-230876), and the CFI-JELF program (38794) for financial support.

**SUPPLEMENTARY**

The PINNs loss function is given as follows:

$$\mathcal{L} = \mathcal{L}_{BC} + \mathcal{L}_{IC} + \mathcal{L}_R + \mathcal{L}_M \quad (S1)$$

$$\mathcal{L}_{BC} = \frac{1}{N_{BC}} \sum_{i=1}^{N_{BC}} \left( \left| -k_1 \frac{\partial T_1^i(t_{\Gamma_1}^i)}{\partial z} \right|_{\Gamma_1} + Q(t_{\Gamma_1}^i)|_{\Gamma_1} \right|^2 + \left| k_1 \frac{\partial T_1^i(t_{\Gamma_2}^i)}{\partial z} \right|_{\Gamma_3} + h\big(T_1^i(t_{\Gamma_2}^i)|_{\Gamma_3} - T_\infty\big) \right|^2 \right) \quad (S2)$$

$$\mathcal{L}_{IC} = \frac{1}{N_{IC}} \sum_{i=1}^{N_{IC}} \left| T_1^i(t_{IC}^i) - T_{ini} \right|^2 \quad (S3)$$

$$\mathcal{L}_R = \frac{1}{N_r} \sum_{i=1}^{N_r} \left| \frac{\partial T_1^i}{\partial t} - \alpha \left( \frac{\partial^2 T_1^i}{\partial x^2} + \frac{\partial^2 T_1^i}{\partial y^2} + \frac{\partial^2 T_1^i}{\partial z^2} \right) \right|^2 \quad (S4)$$

$\mathcal{L}_M$ is defined in Eq. (17).